# Programming Cloud Resource Orchestration Framework: Operations and Research Challenges

Rajiv Ranjan[1,2], Boualem Benatallah[1]

[1] School of Computer Science and Engineering, University of New South Wales
{rajivr, boualem}@cse.unsw.edu.au

[2] Information Engineering Lab, CSIRO ICT Centre, Acton, Canberra
{rajiv.ranjan}@csiro.au

## Abstract

The emergence of cloud computing over the past five years is potentially one of the breakthrough advances in the history of computing. It delivers hardware and software resources as virtualization-enabled services and in which administrators are free from the burden of worrying about the low level implementation or system administration details. Although cloud computing offers considerable opportunities for the users (e.g. application developers, governments, new startups, administrators, consultants, scientists, business analyst, etc.) such as no up-front investment, lowering operating cost, and infinite scalability, it has many unique research challenges that need to be carefully addressed in the future. In this paper, we present a survey on key cloud computing concepts, resource abstractions, and programming operations for orchestrating resources and associated research challenges, wherever applicable.

## 1. Introduction

Cloud computing [86][88][89][90] paradigm is shifting computing from physical hardware- and locally managed software-enabled platforms to virtualized cloud-hosted services. Cloud computing assembles large networks of virtualized services: hardware resources (CPU, storage, and network) and software resources (e.g., databases, message queuing systems, monitoring systems, load-balancers). Cloud providers including Amazon Web Services (AWS), Microsoft Azure, Salesforce.com, Google App Engine, and others give users the option to deploy their application over a network of infinite resource pool with practically no capital investment and with modest operating cost proportional to the actual use.

Key to exploiting the potential of cloud computing is the issue of Resource orchestration. Resource orchestration process spans across range of programming operations, from selection, assembly, and deployment of resources to monitoring their run-time performance statistics (e.g., load, availability, throughput, utilization, latency, etc.) for ensuring consistency and adaptive management. With orchestration, the overall goal is to ensure successful hosting and delivery of applications by meeting the Quality of Service (QoS) objectives of users. QoS is composed of number of functional and non-functional attributes such as performance statistics, consistency, security, integrity, reliability, renting cost, scalability, availability, legal and regulatory concerns. Similar to many recent proposals, we argue that resource orchestration in cloud environments is complicated due to the scale, heterogeneity, and diversity of resource types; and uncertainties of the underlying cloud environment. The uncertainties arise from a number of factors including resource capacity demand (e.g., bandwidth and memory), failures (e.g., failure of a network link), user access pattern (e.g., number of users and location) and lifecycle activities of applications.

In particular, cloud resource orchestration is challenging because applications are composed of multiple, heterogeneous software and hardware resources, which may have integration and interoperation dependencies. Optimal application QoS demands bespoke resource configuration. However, existing literature lacks a detailed, comprehensive cost, performance or feature comparison study of cloud providers. Currently, selection and deployment of cloud resources requires human familiarity with the various providers and extensive manual programming. This is inadequate, given the proliferation of new providers offering resources at different layers. To program a resource orchestrator that can guarantee application QoS fulfillment while automatically coping with the inherent variation in application workload patterns (e.g. request arrival rate, service time distribution, data size, etc.), and the uncertainty (e.g. failure, congestion, overload, energy-consumption, etc.) of cloud environments remains a very challenging problem.

The main contributions of this paper are: (a) advancing the fundamental understanding of key cloud computing concepts and resource abstractions; (b) characterization of cloud resources in a multi-layered stack based on their attributes, granularity, and supported programming operations; and (c) highlighting the main research challenges involved with programming orchestration operations for different cloud resource types.

The remainder of the paper is organized as follows: Discussion on high-level resource orchestration operations required for deploying enterprise applications on clouds is presented in Section 2. Details on major cloud concepts, resource abstractions, and relevant orchestration challenges are discussed in Section 3. Finally, Section 4 summarizes the related work and provides some concluding remarks.

## 2. Abstract Orchestration Operations Required for Hosting Enterprise Applications

The application architecture (e.g., enterprise application, scientific workflows, MapReduce, etc.) determines; how, when, and which orchestration operations should be affected on cloud resources. Though lack of space does not permit discussion about all application architectures, here we discuss orchestration operations for managing enterprise applications in cloud environments. The high-level architecture for enterprise applications is depicted in Fig. 1.

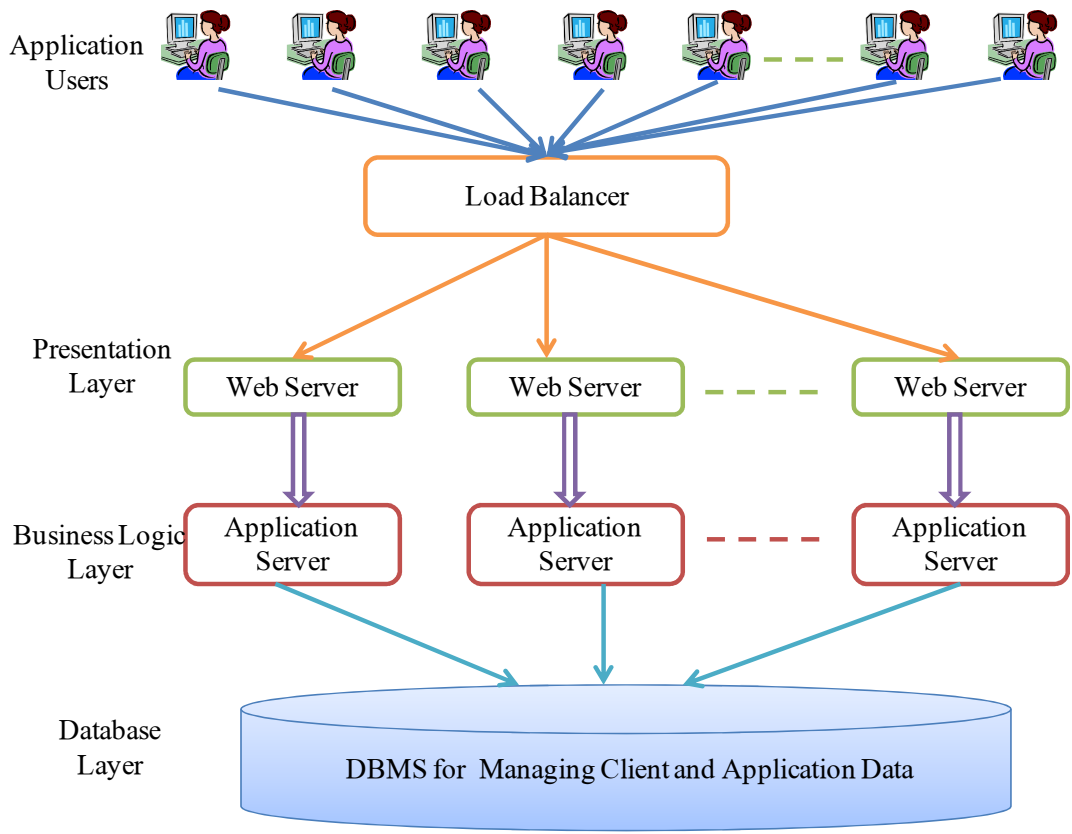

**Figure 1:** High level architecture of a multi-layered enterprise application consisting of clients, load balancer, web servers, application servers, and database management system. The flow of requests between these layers is often complex. Each layer may instantiate multiple software resources (e.g., web server, application server, database server, etc.); each software resource may need to be replicated on multiple hardware resources (e.g., CPU), while load-balancers distribute requests across instances of software resources.

The majority of enterprise applications (e.g. student registration, employee payroll, supply chain management, etc.) are built using multiple tiers to decouple the major functionalities across three software resource layers: i) presentation layer, front-end web servers which are responsible for handling end-user requests and managing state and data for application presentation through rich interfaces (e.g., JSP, HTML, JavaScript, Ajax, etc.); ii) business logic layer, which is hosted within application containers (e.g., Java EE, .NET, etc.) and coordinates application, business entities. It makes logical decisions and evaluations and performs calculations; and iii) data layer, which stores information in relational database server. The information is processed by business entities for decision making. Across the layer, number of orchestration operations need to be programmed to control the resources at design time as well as run time for ensuring the fulfillment of QoS objectives. Briefly stated, these operations are (refer to Fig. 2):

**Resource Selection (both design and run time):** a developer analyzes candidate software resources (such as web server, application server, database server, and load-balancer) to determine whether they can be selected for realizing anticipated application stack and functionality (e.g. supply chain management, wiki, etc.). In addition, the developer needs to verify whether the software resources satisfy the desired functional (e.g. inter-operability with other software resources, compatibility with target hardware resources) and non-functional (e.g. cost, reliability, license management, service level agreements, legal issues etc.) selection criteria.

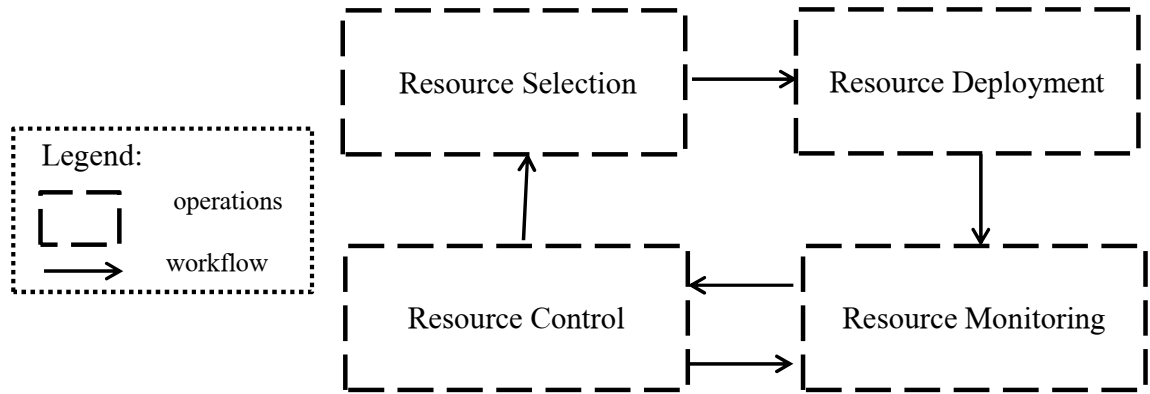

**Figure 2:** Abstract resource orchestration operations in lifecycle of an enterprise application.

Following this, the developer selects hardware resources (CPU, storage, and network) that need to be allocated (capacity planning) to software resources. Such allocation caters for varying levels of QoS requirements that may exist across the layers of enterprise application. For example, a web server that serves images, scripts, and static web pages could be hosted on a CPU resource with moderate attributes (CPU speed, physical memory, CPU cores, and I/O, location, availability zone, etc.). On the other hand, a disk-intensive database server that frequently indexes and updates data should be assigned more powerful CPU resource attributes; otherwise it can become performance bottleneck.

**Resource Deployment (both design time and run time):** Instantiating software resources on hardware resources and configuring them for communication and inter-operation with other software resources. Integration of an application server with the database server is a salient example of this orchestration operation.

**Resource Monitoring (run time):** Monitoring the QoS statistics of deployed hardware resources and hosted software resources in order to detect exceptions. It involves gathering events and information produced by deployed resources; viewing software resource instance QoS statistics, including the number of instances in each state. By analyzing the aforementioned information, a resource orchestrator can detect malfunctions and initiate policy-based corrective actions without disrupting the run-time system.

**Resource Control (run time):** To satisfy QoS objectives of users, set of operations (e.g., scaling-in, scaling-out, data synchronization across replicated database server instances, etc.) must be carried out at run-time for handling the uncertainties (e.g., resource failure, rapid surge in user population, etc.), while ensuring the achievement of QoS objectives as set out in SLA. An example resource control operation could be to upgrade the physical memory size of the existing CPU resource to improve the throughput of the hosted software resource (such as database server). Type and mix of resource control operations will vary depending on the type of cloud hardware and software resource under consideration. For example, data synchronization resource control operation is relevant to a database server, but not to the load-balancer or a web server.

## 3. Cloud Resource Types

Cloud computing follows a service-driven, layered business model. It offers hardware and software resources that can be mapped into three layers (see Fig. 3): Software as a Service (SaaS), Platform as a Service (PaaS), and Infrastructure as a Service (IaaS). Hardware resources along with software resources form the basis for delivering IaaS and PaaS. The top layer focuses on application services (SaaS) by making use of those services provided by the lower layers. PaaS/SaaS services are often developed and provided by 3rd party service providers, who are different from the IaaS providers. We explain each resource type through examples and analyze the core research challenges involved with programming orchestration operations.

**(1) Infrastructure as a Service**

The CPU, storage, and network resource in cloud environment is supplied by a collection of data centers that are installed with hundreds to thousands of physical resources [87] such as cloud servers, storage repositories, and network backbone. These resources expose certain attributes (see Table I) that can be allocated to software resources. *In general, attributes define consumable features and functions that are available from hardware resources.* Providers manage these physical resources through hardware virtualization technologies such as Xen [16], Citrix, KVM (open source), VMWare [17], Microsoft, etc.

Virtualization allows providers to get more out of physical resources by allowing multiple instances of virtual cloud resources to run concurrently. Each virtual resource believes it has its own hardware. Virtualization isolates the hardware resources from each other, thereby making fault tolerant and isolated security context behavior possible. It enables more efficient utilization by providing the flexibility, agility and scalability needed for a physical resource to support multi-tenancy. Cloud computing is inclusive of virtualization and a way to implement it. However, clouds can also be implemented without virtualization as well. For example, Google, EmuLab, and iLO offer hardware resources without virtualization to manage their physical resources. We list the major providers at IaaS layer in Table II.

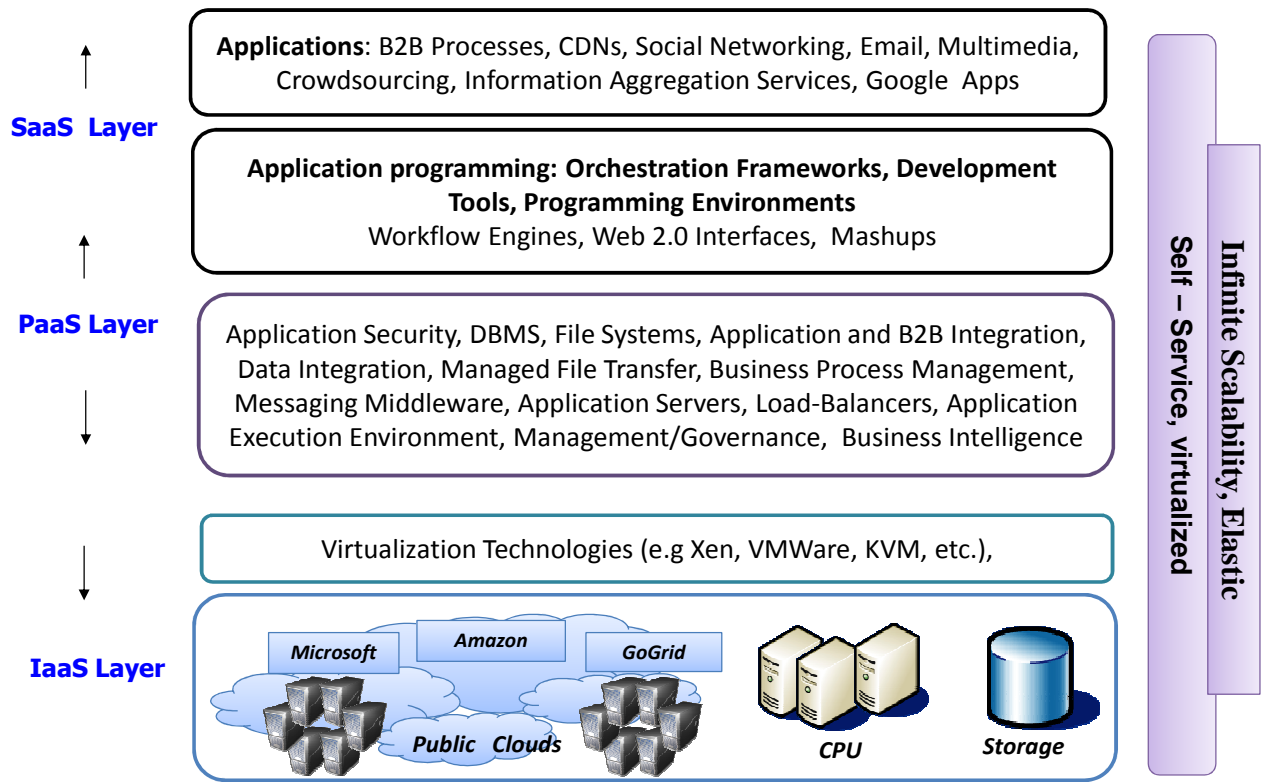

**Figure 3:** Reference cloud resource stack. The architecture provides a layered approach to characterizing resources based on their attributes and granularity.

A CPU resource [37] is essentially is a piece of virtualization software running on the physical CPU resource. It is the most common method of exposing the computational power to software resources; where one gets finer-granularity accessibility and flexibility at super-user level and that can be used to customize the placement of software resources for QoS. It emulates the properties of a physical CPU resource by providing a virtual processing unit (CPU), network card, physical memory, hard disk, keyboard, and so forth. A physical CPU resource can host multiple virtual CPU resources. Each CPU resource can be a bare metal (in an infrastructure-as-a-service provider or a sandbox environment (in a platform-as-a-service provider, such as Google App Engine). For example, AWS EC2 offers 11 CPU resource types (e.g. Small, Large, Extra Large, Micro, and others) that have variable cost/performance efficiencies and are managed by Xen virtualization technology.

| Hardware Resources | Attributes | Supported Orchestration Operations |
|---|---|---|
| CPU | cores, speed, family, physical memory capacity, storage capacity, addressing bits, input/output performance, renting cost, type (single or cluster of templates), resource sharing (multi-tenant or dedicated), physical location of cloud, availability zone, QoS SLA | start, stop, restart, select, mount storage, monitor, reconfigure, assign IP, select cloud location, select availability zone, scale-in, scale-out |
| BLOB Storage | type (persistent or non-persistent), storage size, storage format, renting cost, location of host cloud, availability zone, QoS SLA | create new volumes, attach volumes to CPU resource, create new buckets, upload file, download file, scale-in, scale-out, monitor |
| Network | IP Type (static or dynamic), version (IP V4 or IP V6), renting cost, message encryption cost, URL, data transfer in cost, data transfer out cost, connection-hour, QoS SLA | allocation of IP addresses, URL, ports, availability zone, and VPN to CPU resources, monitor |

**Table I.** Hardware resource types, their attributes, and list of supported orchestration operations.

Users can perform number of orchestration operations on CPU resources such as starting an instance of CPU resource, assigning IP address to CPU resource, etc. Operations relevant to CPU resources are shown in Table I. In reality, multiple CPU resources can be instantiated on a single cloud server, wherein they have dedicated access to physical resources such as processing power, memory, and local storage. However, other physical resources including the network and the disk subsystem are not dedicated but shared among CPU resource instances.

The second IaaS-level hardware resources is the data storage resources that allows users to store raw application data on virtualized disks and access them anytime from any point on the Internet. They are also referred to as the Binary Large Object (BLOB) for binary files storage resources such as Amazon S3, Azure Blob, Google Cloud storage, etc. These storage resources are different from the local storage (for example, the local hard drive) in each CPU resource, which is temporary or non-persistent and cannot be directly accessed by other instances of CPU resources. BLOB storage resources can hold video, audio, photos, archived email messages, or anything else, and they let applications store and access data in a very flexible way. They aim to enforce fault-tolerant behavior through redundancy and intelligent distribution of data.

Recall that a CPU resource has access to its local hard drive; however, by default, local drive is non-persistent and hence once the instance of a CPU resource is terminated, its local storage contents are purged. To overcome this issue, providers including AWS EC2 and Microsoft Azure offer off-instance storage resources that persist independently from the life of a CPU resource. These off-instance storage resources are referred to as the Elastic Block Store (EBS) and XDrive in case of AWS EC2 and Microsoft Azure respectively. They are particularly suited for application that requires database, file system, or access to raw level storage. Principal advantages of architecting application using off-instance storage include: i) data is automatically replicated, this prevents data loss due to failure of any single hardware component and ii) one could create point-in-time snapshots of data and back it to the cloud specific unstructured, highly-available, storage resources (such as Amazon S3 and Azure Blob).

As the need for high volume data transfer and communication across network boundaries grows for applications, networking resources (e.g. routers, switches, communication bandwidth, AWS elastic IP, OpenFlow, AWS security group, etc.) become a vital component at the IaaS level. Network resources provide variety of functionalities including bandwidth, virtual overlays for isolating traffic, guarantying message delivery delay, encrypting communication channels, and network monitoring. Orchestration operations relevant to network resources include allocation of IP addresses, URL, ports, availability zone, and VPN to CPU resources. While all CPU and storage resource providers basic network resources, Table II shows the list of providers who offer specialized resources such as firewall, network virtualization technologies etc.

| Hardware Resources | Major Providers |
|---|---|
| CPU | Amazon EC2, AppNexus, BlueLock, FlexiScale, GoGrid, Joyent, Rackspace, Terremark, EmuLab, HP iLO, Microsoft Azure, IBM BlueCloud, Rimuhosting, RHEV-M, PrimaCloud, Hosting365, Engine Yard, Savvis, ElasticStack, Enki, AT&T Synaptic Cloud, CloudSigma, ElasticHosts, FiberCloud, LayeredTech, Locaweb Cloud Server Pro, Maxnet, Navisite, SunGard, Verizon |
| BLOB Storage | Amazon S3, Microsoft Azure Blob, GoGrid cloud Storage, Mediamax, EMC, HP, IBM, ParaScale, NetApp, Flexiant, Nirvanix, Joyent BingoDisk, 10gen, Rackspace cloud storage, Google Big Table, Google File System, Amazon EBS, Aprigo NINJA, EMC Atmos Online, Zetta, Backupify, BKol, Mozy, Jungledisk, Online Backup, SpiderOak, Zmanada Cloud Backup |
| Network | CISCO, OpenFLow, Citrix, IBM |

Table II. Major providers of hardware resources.

### (1.1) Research Issues in Programming Orchestration Operations at IaaS Layer

Summary of research issues relevant to different IaaS resource types is depicted in Table III. Detailed discussion on these research issues follows next.

**Optimal Selection and Comparison.** The diversity of offering at this layer leads to a practical question: how well does a cloud provider perform compared to the other providers? For example, how does a user compare the cost/performance features of CPU, storage, and network resources offered by AWS EC2, Microsoft Azure, GoGrid, FelxiScale, TerreMark, and RackSpace. Answering this question can benefit both users and providers. For a user, the answer can help in selecting the best size and mix of hardware resources to ensure that the application meets its QoS targets. For instance, a low-end CPU resource of Microsoft Azure is 30% more expensive than the comparable AWS EC2 CPU resource, but it can process application workload twice as quickly. Similarly, a user may choose one provider for storage intensive applications and another for computation intensive applications. On the other hand, providers can use the answer to identify their areas of competitive disadvantage and thus lead to redesign and improvement. For instance, a provider should allocate more resources into optimizing their storage if the performance significantly lags behind competitors.

**Avoiding Deadlock.** Since, the state transitions in hardware resource are expensive, irreversible, and time-consuming, therefore orchestration operations have to be methodologically implemented and executed. For example, it could take several minutes to boot a high-end CPU resource in AWS EC2. Over the lifecycle of an application, an orchestrator has to realize multiple operations (see Table I), in a sequenced manner, across hardware resources that may belong to geographically distributed clouds. Orchestration operation on particular class of hardware resource (e.g. a CPU resource) is enforced by invoking their respective (provider-specific) web-service API [24][25]. Further, these operations are significantly complex and thus prone to failures. For example, in a multi-step orchestration operation of allocating a CPU resource to a software resource followed by assigning an elastic network IP and mounting an EBS resource; if one of the immediate operations fails or throws unexpected error, a trivial implementation would fail stop, leaving the system in inconsistent state. Ensuring dead-lock free orchestration environment for dealing with high level of concurrency and network traffic arising from potentially large number of over-lapping, distributed orchestration requests remains an open research issue.

**Dynamic Adaptation.** The failure or congestion of network links or malicious external interference that changes the state of hardware resources are sometimes inevitable, given the scale, dynamics, and complexity in cloud computing, the crash or malfunction of a hardware resource, changes in application workload patterns, or overloading of a hardware resource are common phenomenon. The recent very high-profile crash of AWS EC2 cloud, which took down the applications of many SMEs, is a salient example of unpredictability in cloud environments. Theoretically, the elasticity provided by cloud computing can accommodate even unexpected changes in capacity, adding hardware resources when needed, and reducing them during periods of low demand, but the orchestration decisions to adjust capacity must be made frequently, automatically, and accurately to be cost effective. To develop orchestration techniques that can dynamically predict and capture the relationship between application QoS targets, current hardware resource allocation and changes in workload patterns, in order to adjust resource allocation remains an active and open area of research.

**Energy Efficient Allocation.** In recent years, energy efficient allocation of resources such as CPU, memory, storage, and network to applications have emerged as one of the critical requirement because of focus on minimizing carbon footprint and electrical power bills. Efforts have focused on the fabrication of energy-efficient hardware, such as low power energy efficient CPUs, low power computer monitors, and solid state drives to mitigate energy consumption. Use of software based approaches, such as resource allocation and task consolidation, to minimize energy consumption has also received certain attention from research community. The optimal assignment of hardware resources to the software resources of an application in an energy efficient manner remains a hard and open research problem.

**Data Security and Privacy.** BLOB storage resources (e.g., Amazon S3, Azure Blob, SkyDrive, etc.) are not secure [78] by nature. Hence, application data stored over these storage resources are inherently under the risk of: (a) data exposure (confidentiality); (b) data tampering (integrity); and (c) denial of access to data (availability). The instance of accidentally or intentionally (by the cloud provider) exposing an application data to the third party applications on the cloud is known as the data confidentiality risk. Next, data integrity risk is defined as the situations that arise due to tampering of data by the malicious third parties or the cloud provider. Therefore, the application data must be accurately encrypted not only while being transmitted over network links but also while at rest on storage resources On the other hand data availability refers to the risk of denial of access to application data by third parties on the cloud or by the cloud provider itself.

| Resource Name | Research Issues in Programming Orchestration Operations |
|---|---|
| CPU | optimal selection and comparison, avoiding deadlock, dynamic adaptation, energy efficient allocation, inter-operability |
| BLOB Storage | optimal selection and comparison, avoiding deadlock, dynamic adaptation, energy efficient allocation, data security and privacy, scalable data indexing, inter-operability |
| Network | avoiding deadlock, dynamic adaptation, energy efficient allocation, data security and privacy, inter-operability |

Table III. Classifying major research issues involved with programming resources at IaaS layer.

**Scalable Data Indexing.** Though, BLOB storage resources can hold video, audio, photos, archived email messages, or anything else, and they do not expose any API for data indexing. With the increase in the scale and the size of data (e.g., PetaBytes), efficient indexing and distribution becomes a critical issue. The challenge is further aggravated in case of live (e.g., live video streaming) and interactive data, where size of distribution (hence the indexing complexity) in not known in advance. To support efficient data storage and indexing on the scale of PetaBytes, it is mandatory to design indexing algorithm to enable access and search over BLOB. It is worth noting that none of the existing BLOB resources exposes *data indexing APIs*, it is upto the application designer to come-up with efficient indexing structure that can scale to large data sizes.

**Inter-operability.** To improve resilience to uncertainties, an intuitive solution can be to deploy applications across multiple IaaS providers. Unfortunately, most of the existing providers are not compatible with each other. They tend to have proprietary APIs, which are not explicitly designed for cross-cloud interoperability. To tackle such heterogeneities, there is a requirement to enforce standardization across layers of cloud resource stack. Recent developments including Simple Cloud [24],

Delta Cloud [25], JCloud [69], and Dasein Cloud [70] simplify this task by implementing single API that abstracts APIs related to multiple clouds such as AWS EC2, GoGrid, etc. Though aforementioned APIs can simplify implementation of multi-cloud resource orchestrator, system designers still need to cater for the heterogeneities that prevail in terms of virtualization technologies, resource naming, and cost/performance efficiencies, etc.

(2) Platform as a Service

This layer features a rich pool of software resources (see Table IV) including orchestration frameworks, development tools, programming environments, and appliances that facilitate the end-to-end life cycle of developing, testing, deploying, and hosting applications. Following software resource categories are relevant at this layer.

**Development Tools and Programming Environments:** They provide a range of programming languages and developments tools ranging from salerforce.com's custom language, APEX; to Microsoft Azure's C#; to Java; to Python and Ruby. All of these can be used for developing cloud-ready applications. Web 2.0 Interfaces (Ajax, IBM Workplace) that help developers in creating rich, cost-effective user-interfaces for browser-based applications are also part of PaaS layer.

| Software Resource | Provider | Description |
| --- | --- | --- |
| Django | Google | rapid development and deployment of python-based web applications on Google App Engine. |
| Hadoop | Apache Open Source Foundation | aids in programming data intensive applications |
| .NET SDK | Microsoft | aids in development and runtime management of applications. SDKs also available in Java and Ruby. |
| APEX | Salesforce | development of applications over Salesforce.com |
| Web 2.0 | Ajax, IBM workplace | creating rich, cost-effective user-interfaces |
| Rational Software | IBM | software delivery service in a cloud environment |
| LoadStorm | LoadStorm.com | load and performance testing tool for cloud resources |
| BrowserMob | BrowserMob.com | load-testing tool for cloud resources |
| Ruby on Rails | Heroku, Engine Yard | developing ruby on rails application over Heroku and Engine Yard clouds |
| EdgePlatform | Akamai | Content, site, application delivery |
| Zoho Creator | Zoho | toolkit to build and deploy business applications |

Table IV. Major providers for development tools and programming environments.

Recently, MapReduce has gained increased attention as a programming environment for processing and analyzing large data sets (e.g., terabytes, petabytes). The run-time system of MapReduce takes care of technical complexities related to partitioning the input data, scheduling the application's execution across a set of machines, handling machine failures, and managing the required inter-node communication Originally, it was introduced by Google for processing and generating large data sets on distributed computing infrastructures (Google's clusters). Hadoop, an open source implementation of MapReduce, has been successfully implemented over AWS EC2 (available via AWS Elastic MapReduce Orchestrator) and the Yahoo cloud Supercomputing Clusters. For example, New York Times utilized Hadoop over Amazon EC2 for converting its news content from 1852 to 2002 to PDF. They did so in less than 24 hours with 100 AWS CPU resources and some helper scripts. Officially, Google App Engine does not support MapReduce programming environment, but in the recent past certain open source projects [75] have started implementing subset of MapReduce API for it. On the other hand, one can use Dyrad (which is still in Beta release stage) for implementing MapReduce over Microsoft Azure.

Other offerings at this layer have also focused on providing a complete, sandboxed environment for developing enterprise applications. Django from Google App Engine aids in rapid development and deployment of python-based enterprise applications. However, due to lack of configurability, Django cannot be used for developing legacy applications, nor can it use for enterprise applications built upon relational database technology. AppScale, which is an open source implementation of Google App Engine, enables execution of applications on local clusters with possibility to scale-out to external clouds (such as AWS EC2 and Eucalyptus). Heroku (http://www.heroku.com) and Engine Yard provide a programming and deployment environment for Ruby on Rails based enterprise applications. Microsoft Azure offers a wide range of alternative programming environments such as Azure SDK, workflow management service, and access to SQL data stores for building enterprise applications.

**Appliances:** An appliance is pre-configured, self-contained, virtualization-enabled, and pre-built software resource unit (database, web server, application server, load-balancers, etc.) that can be integrated with other compatible appliances for architecting complex applications. Primarily it is the goal of resource orchestrator to select, assemble, deploy, and manage a set of appliances delivering particular application functionality. There is currently no widely accepted standard for appliance virtualization format, which means that an appliance is built for a specific virtualization technology and possibly will not run on cloud resources that are managed by non-overlapping virtualization technologies. For example, a VMware database appliance cannot be deployed on AWS EC2 CPU resource, as AWS only supports appliances in Xen virtualization format (also referred to as Amazon Machine Image virtualization format). Similar to hardware resources at IaaS layer, appliances have basic attributes and they support number of orchestration operations (see Table V).

| Attributes | Supported Orchestration Operations |
|---|---|
| feature (web server, database server, load-balancer, authorization server, etc.), virtualization format (Xen, VMware, etc.), environment (host operating system, implementation language such as Java, .Net, PHP, Ruby on Rails, etc.), legal and regulatory issues, security, reliability, licensing terms and costs, initialization scripts | select, allocate hardware resources , integrate with other appliances, install script, monitor, create, migrate, scale-in, scale-out, log-in, log-out, install software, replicate, synchronize, backup, delete |

Table V. Attributes and orchestration operation relevant to an appliance.

There have been some standardization efforts in this space such as Open Virtualization Forum (OVF). OVF [39] describes an open, secure, portable, efficient, and generic format for the packaging and distribution of appliances. Though cloud vendors including Microsoft, IBM, Dell, HP, VMWare, and Xen have supported OVF initiative, its widespread adoption is still questionable. Hence, the problem of heterogeneity in packaging appliances continues to be an obstacle. To overcome this heterogeneity, providers including Amazon EC2, RightScale, and CloudSwitch offer custom scripts that can be utilized to port [41] appliances from one virtualization format to another.

Most cloud providers host a private repository [27] of appliances that can be orchestrated either through command line or web-portal interfaces. VMware [29] offers one of the largest marketplaces of appliances. Bitnami [22][23], 3Tera [10], CloudMarkets [30] , and rPath [28] are other major players in this space. The type and mix of appliances varies across providers. For example, GoGrid's appliance repository [40] consists of in-house (packaged by GoGrid) appliances as well as $3^{rd}$ party appliances from other providers such as Bitnami, Gigaspaces, cPanel, cloudKick, etc. Refer to Table VI for detailed insight on appliance providers.

Here, we also distinguish between *basic appliances and composite appliances*. A basic appliance delivers single abstract functionality that may not be sufficient to architect a fully functional application. Example includes a web server appliance, a database appliance, a monitoring appliance etc. In this approach multiple basic appliances need to be integrated to create a functional application. 3Tera Applogic adopts this approach as regards to composing appliances. In contrast, a composite appliance encapsulates number of software resource units to support a standalone, fully functional application. For instance, Bitnami's Redmine composite appliance encapsulates multiple software resource units including MySQL and Ruby on Rails to deliver web-based project management functionality. Notably, multiple composite appliances can also be integrated to create more sophisticated applications. For example, Integration of Bitnami's Redmine and Subversion appliances to create a fully functional project management application.

| Appliance Provider | Categories | Virtualization Technology | Description |
|---|---|---|---|
| Bitnami | basic, composite, customizable | AMI, VMWare ESX, Native | provides ready to deploy appliances created from open source software such as Apache, MySQL, PHP, Ruby, etc. |
| RightScale | basic, composite, customizable | AMI, Microsoft Hyper-V, VMWare ESX | offers more than 40,000 RightScale and third party appliances |
| rPath | basic, composite, customizable | VMWare ESX/ESXi, vSphere, vCloud, Citrix XenServer, Microsoft Hyper-V, KVM, Virtual Iron, AMI, Native | repository of appliances that can be deployed on physical (native), virtualized (VMWare, Xen, etc.), and cloud environments (Amazon EC2, Eucalyptus, OpenStack, Rackspace, GoGrid, IBM BlueLock, Globus, etc.) |
| CloudKick | basic, non-customizable | builds upon Libcloud API to abstract interaction with multiple clouds | a centralized monitoring appliance for multiple cloud providers including Amazon EC2, GoGrid, Rackspace, Joyent, etc. |
| Amazon | basic, composite, customizable, non-customizable | AMI | one of the pioneers, in addition to basic/composite appliances they provide number of non-customizable (EC2 load-balancer, CloudWatch, SNS, and SQS) appliances that can be directly instantiated on EC2. |
| 3Tera | basic, customizable | Xen, Microsoft Hyper-V, VMWare ESX | provides a catalog of basic appliances that can be integrated and deployed via a drag and drop interface |
| Microsoft Azure | basic, customizable, non-customizable | Microsoft Hyper-V | developers of cloud applications can potentially mix and match number of basic/customizable (e.g. SQL Azure) and non-customizable (e.g. Service Bus, and Access Control) appliances |
| CloudMarkets | N.A. | N.A. | a catalog of most popular Amazon specific appliances. Currently indexes 15000+ AMI appliances of Bitnami, Flurdy, Turnkey, mental images, canonical, zend, JumpBox, etc. |

Table VI: Major Appliance providers and their technical details.

At this layer we also distinguish between *customizable appliances and non-customizable appliances*. Bitnami, Turnkey, and rPath, offer appliances that can be customized in terms of their mapping to hardware resources. For instance, a Bitnami AMI (Amazon Machine Image) appliance can be mapped to one of the specific CPU resource type depending on anticipated QoS targets. Similarly, with customizable appliances users have flexibility to mount EBS volumes, if persistence of application data is a requirement.

On the other hand, providers such as AWS EC2 and Microsoft Azure offer non-customizable appliances that can directly (without any further modification) be integrated into an application. For instance, AWS offers load-balancer and a monitoring appliance (AWS Cloudwatch) for integration with applications to be hosted on AWS EC2. Similarly, users planning to host applications on Microsoft Azure cloud can leverage Service Bus appliance for common message and communication methods such as events, one-way messages, remote procedure calls and tunnels for streamed data. These appliances are pre-configured, pre-installed, and completely managed by the providers; users have no flexibility as regards to changing their hardware resource allocation.

Scalable Database Management Systems (DBMS) are critical for both update intensive and decision support application workload. There are two types of database appliances one can use for realizing database layer. They include [76][77]: SQL appliance and NoSQL appliance. SQL appliances are the traditional relational database systems (e.g., MySQL, SQL Server, PostGres, Oracle, etc.) that are pre-configured and pre-installed within a virtualized container (e.g., AMI, Xen, KVM, etc.). In SQL appliances, the access pattern of data (e.g., complex join queries) are not known in advance, hence assumptions are made regarding the access patterns and related query load. The data is stored in tables that have fixed schema. Structured Query Language (SQL) is used as the generic language that allows operation (e.g., insert, delete, update, etc.) over the data. Fundamentally, relational databases have proven to be good at managing structured data, especially in the application scenario where transactional integrity (ACDI properties) is a requirement. SQL appliances support only planned scalability rather than dynamic scalability, which something users (application engineers) cater for in the cloud hosted applications. SQL appliance continues to be mainstay for large, complex enterprise applications.

NoSQL appliance [76][77] has recently emerged to complement traditional relational database systems (SQL appliance). Unlike SQL appliances, they do not have support for ACID transaction principles; rather offer weaker consistency properties, for example eventual consistency. NoSQL appliances allow data access based on predefined access primitives such as key-value pair; given the exact key, the value is returned. This well defined data access pattern results in better scalability and performance predictability as compared as SQL appliances. Further, there are four categories of NoSQL appliances including Key-value sores (e.g., Amazon Dynamo), BigTable Clones (e.g., HBase, Cassandra, etc.), Document Databases (e.g., CouchDB, MongoDB, etc.), and Graph Databases (e.g., Neo4j, AllegroGraph, etc.). Unlike SQL appliances, the NoSQL appliances: (i) do not require fixed table schemas; (ii) do not support foreign key relationships and join operations; and (iii) support horizontal scalability. NoSQL appliances trade ACID-compliant data practices of SQL appliances for important advantages in schema flexibility, predictable elasticity, massive availability, and simplified operational manageability.

**Orchestration Frameworks:** Orchestration frameworks provide a thin software layer that hides the nuances of the underlying virtualization tiers, hardware resource abstractions, and appliances. They expose the cloud resources to users for on-demand manipulation through web service APIs, command line tools, or web portals. Providers with notable presence in this space include Amazon (though BeanStalk and Cloudformation offerings), CloudSwitch, CA (through acquisition of 3Tera), rPath, Engine Yard, Elastra, RightScale, VMware vCloud Director, Eucalyptus, Flexiant, Enamoly, Microsoft Azure Platform Appliance and Fabric Controller, Platform ISF, and VMOps. For example, RightScale orchestrator offers an framework that helps users to manage their applications over AWS EC2 and Rackspace. It provides web-based interfaces for deploying CPU resources, selecting appliances, integrating appliances, configuring automated storage backups, etc. On the other hand, Amazon Beanstalk Orchestrator aids in creation, deployment, and management of multi-layered enterprise application by deploying Amazon specific appliances and $3^{rd}$ party development (java tomcat container) tool over AWS hardware resources.

### (2.1) Research Issues in Programming Orchestration Operations at PaaS Layer

**Optimal Selection.** Overall, performing orchestration at this layer is challenging due to the complexities discussed next. Appliance selection operation requires comprehensive understanding about the technical details, capabilities, and interoperation ability of competing appliances. In particular, the orchestrator needs to evaluate whether an appliance can deliver the requested functionality (e.g. database server, source code management server, etc.). If a group of appliance is going to be selected, then they have to meet integration (inter-operational ability) constraints. Finally, the compatibility of an appliance with the virtualization technology of the target cloud has to be considered during the selection process. Other important selection criteria may include appliance's host operating system and its programming environment (e.g. Java, .Net, PHP, Ruby on Rails, etc.).

In addition, there are non-functional criteria that further complicate the selection process. They include the cost of licensing (open source vs. licensed software), reliability, legal and regulatory issues. For example, certain U.S.A export laws prohibit use of specific type of cryptographic technology by a software resource unit, which may be deployed within its geographical jurisdiction. Hence, orchestrator needs to consider such non-functional correlation between an appliance's attributes and location of anticipated cloud location. Overall, given the set of functional and non-functional attributes with single or multiple optimization constraints and large set of available offerings (appliances), the selection problem has been shown to be a NP-complete class of computational problem.

**Automated Appliance Integration.** Mapping an application (set of appliances) to hardware resources requires execution of a complicated sequence of calls to cloud APIs [24][25] to search, load, and control resources and to configure cloud specific isolation, security, and communication environments. An application can be composed of several appliances. The deployment procedures and order of their executions are unique to each application and cloud environment. Dependencies between various appliances in an application must be taken into account to ensure correct deployments. Most of the available IaaS [38][18][15][36] and PaaS [41][17][24][25] clouds provide APIs to deal with single CPU or storage resource orchestration primitives, lacking facility for treating application as a single entity and handling the dependencies among different appliances; for example, dependencies between CPU resources hosting appliances for different tiers (web, application, database, and load-balancer tiers in case of web applications) of an enterprise application.

**Business Process Integration.** For many SMEs, there is a large amount of IT assets that are in house, in the form of line of business applications that are unlikely to ever be migrated to the cloud. Further, there is huge amount of sensitive data in an enterprise, which is unlikely to migrate to the cloud due to privacy and security issues. As a result, there is a need to look into issues related to integration and interoperability between the software resources on premises and in the cloud. (i) Data Management: not all data will be stored in a relational database in the cloud, eventual consistency (BASE) is taking over from the traditional ACID transaction guarantees, in order to ensure sharable data structures that achieve high scalability and (ii) Business process orchestration: how does integration at a business process level happen across the software on premises and application in the cloud boundary? Where do we store business rules that govern the business process orchestration?

**Comprehensive Monitoring.** Monitoring activity involves dynamically tracking the QoS parameters related to hardware and software resources (appliances), the physical resources they share, and the applications running on them or data hosted on them. Monitoring tools can help an administrator or developer as regards to: (i) keeping their resources and applications operating at peak efficiency; (ii) detecting variations in resource and application performance; (iii) accounting the SLA violations of certain QoS parameters; and (iv) tracking the leave and join operations of resources due to failures and other dynamic configuration changes.

Existing monitoring appliances such as CloudWatch, Cloudrack, Azure Fabric Controller have following limitations including: (i) they do not have ability to monitor across multiple cloud providers; (ii) they are tailored towards monitoring only hardware resources (such as CPU) with support for limited QoS parameters; (iii) they do not support monitoring for individual software resources (e.g., web server, application server, database server, etc.). For example, CloudWatch is not capable of monitoring information related to load, availability, and throughout of each core of CPU resource and its effect on the QoS (e.g., latency, availability, etc.) delivered by the hosted software resource (e.g., J2EE application server or MySQL database server). Hence, there exists a considerable research issue in developing a monitoring appliance that can: (i) monitor both hardware and software resources; (ii) support a comprehensive list of QoS parameters for each resource type; and (iii) reason about the inter-dependencies of QoS parameters across the hardware and software resource layers.

**BIG DATA Management.** In last three decades many research groups have focused on large scale data management in traditional enterprise setting. However, cloud computing and its available NoSQL appliances (e.g., AWS SimpleDB, MongoDB, Azure Table Storage, etc.) and SQL appliances (e.g., SQL Azure, Amazon MySQL, etc.) has its own research challenges as regards to programming orchestration operations (e.g., selection, scale-in, scale-out, synchronize, replicate, backup, etc.). Supporting adhoc querying in top of NoSQL stores and providing hard data consistency guarantees remains an open research problem. Further, it is not clear how NoSQL store will perform for different classes of applications (e.g., enterprise, eResearch, etc.) and workload (e.g., decision support, I/O intensive, etc.). Developing techniques, that can augment cloud-based load-balancing appliance (e.g. AWS Elastic Load Balancer) with data workload characterization intelligence (density and distribution of data; composition of queries) for improving the QoS (i.e. query latency and database service throughput) remains a popular research topic.

**Simplified Resource Abstractions.** While hardware resources and software resources (aka. appliances) continue to be the main resource abstraction in clouds, more sophisticated resource abstractions are emerging including realization of cluster computing on cloud-based hardware resources to enable high-performance computing; integration of monitoring, load-balancing, and auto-scaling appliances to handle surge in application traffic; etc. These more advanced resource abstractions share all the operational complexities with the more fundamental resources. Devising a simplified cloud resource abstraction that can be presented to non-technical users and that thrives for achieving the right balance between the exposing and hiding of information remains an open research problem. This is an aspect [42] which is not very well handled by the existing orchestration frameworks.

### (3) Software as a Service

All the applications that run over the SaaS provider managed cloud and provide a direct service to the end-users are part of the Software as a Service (SaaS) layer. A SaaS provider may own or rent the underlying PaaS and IaaS-level resources. SaaS applications are accessed over the Internet and typically charged on a subscription basis. In our classification, SaaS is at the top of cloud reference stack and provide users with ready-to-use applications. SaaS is a high level clustered service abstraction that logically integrates multiple, inter-operable PaaS and IaaS level cloud resources to deliver specific kinds of application functionalities (e.g. social networks, sales management, project management, security solutions, personal productivity, etc.) to end-users. The selection, assembly, deployment and run-time management of cloud resources are completely invisible to users, as these orchestration operations are handled behind the scene by SaaS providers. Although users are not allowed to change the basic configuration of SaaS such as allocation of hardware resources or integration of new software resources to existing application, they can customize the SaaS at user-interface level for specific needs. While the offerings at this layer are still emerging, we discuss some of the popular ones in Table VII.

At this layer, we distinguish between pay-per-use and free SaaS applications. Example of pay-per-use offerings include SalesForce.com (http://www.salesforce.com) and Clarizen.com (http://www.clarizen.com), which expose CRM (Customer Relationship Management) and project management applications respectively. NetSuite (http://www.netsuite.com) offers a fully-integrated financials, accounting, CRM, inventory, and ecommerce applications. Appirio (http://www.appirio.com) provides an integrated application suite that covers entire spectrum of business processes in enterprises from project management to resource planning. By far, the uniqueness of Appirio is its ability to deploy SaaS over the cloud resources exposed by Amazon EC2, SalesForce.com, Google AppEngine, and Facebook.

Other examples of free SaaS applications at these layers are Google Mail, Google Documents, Google Maps, and Google Calendar. In this part, we also consider some arbitrary applications, such as YouTube, that are created and managed through end-user intelligence (such as uploading a particular video stream). Social networking media such as Facebook, Wordpress, MySpace, LinkedIn, and Twitter are also some of the examples of free online SaaS applications that have gained significant popularity in last five years.

This layer also features certain SaaS applications that rely on massive scale aggregation and extraction of information from crowds of people. For instance, Amazon's Mechanical Turk (MTurk) provides on-demand access to task forces for micro-tasks such as image recognition, language translation, etc. Several organizations including DARPA and various world health and relief agencies are using platforms such as MTurk and Ushahidi to crowd-source information through multiple channels, including SMS, email, Twitter and the Web in general. Some crowdsourcing SaaS such as Iowa Electronic Markets and Digg are more controlled and are targeted at predicting events or promoting popular ideas.

| SaaS | Category | Provider | Functionality |
|---|---|---|---|
| SalesForce.com | pay-per-use | SalesForce | Customer Relationship Management |
| NetSuite | pay-per-use | NetSuite | CRM, accounting, inventory, ecommerce |
| Clarizen.com | pay-per-use | Clarizen | Project management systems |
| Appirio.com | pay-per-use | Appirio | Business process modeling, integration, data migration, training/change management service, SaaS customization, ROI/TCO analysis, Prototyping, complete suite of enterprise applications (sales, marketing, support, finance, HR, collaboration) |
| PingConnect On-Demand SSO for SaaS | pay-per-use | pingidentity | Eliminates redundant administrative work and prevents unauthorized access by automating Internet user account management for every major SaaS application include SalesForce.com, Google Apps, etc. |
| Aria | pay-per-use | Ariasystems | On-demand billing and invoicing system. It can also handle payment processing, subscription management, metered usage plan, end-user analytics, and marketing |
| eVapt | pay-per-use | eVapt | Provides billing and subscription management services for SaaS, PaaS, and IaaS |
| Ltech | pay-per-use | Ltech | Cloud consulting services |
| OpenID | Free | OpenID Foundation | Provisions single digital identity across the Internet |
| Save My Table | pay-per-use | Savemytable | Provides a web-based subscription service that empowers a restaurant's host stand |
| Office Live | pay-per-use | Microsoft | Let end-users store and share documents on-line in a collaborative workspace. Office Live is integrated with Microsoft's Office suite |
| Google Apps | Free | Google | Google Docs – Online office suite<br>Google Maps – map-based services<br>OpenSocial – a common API for developing social applications across multiple websites |
| Acrobat.com | pay-per use | Acrobat | Set of applications for on-line collaboration including file sharing in workspaces, web conferencing, and on-line storage |
| Dropbox | free/pay-per-use | Getdropbox | A cloud based file storage, file sharing and synchronization service |
| Egnyte | Pay-per-use | Egnyte | On-demand file system for enterprises. Features include file sharing, file versioning, automatic backups, locking, and automatic update |
| Mechanical Turk | pay-per-use | Amazon | scalable crowdsourcing workforce |
| Digg.com | Free | Digg | news aggregation |
| Youtube | Free | Youtube | video portal |
| Facebook | Free | Facebook | social networking site, development tools and execution environment for creating and customizing social networking applications |
| Iowa Electronic Markets | Free | The University of Iowa | For predicting outcomes of political races |

Table VII: Major SaaS Applications and their details.

## (3.1) Research Issues in Programming Orchestration Operations at SaaS Layer

**License and Users Management.** For SaaS providers, managing licensing has become a major orchestration issue. In effect, hosting an application over cloud would make proprietary software accessible to millions. This raises a serious challenge for SaaS providers, who need to respect the rules and regulations of open source technologies and licensed software in clouds, yet making them interoperable. SaaS providers need to orchestrate the management of user accounts for billing and accounting purposes.

**Optimal Selection.** Similar, to resources at PaaS and IaaS layer, an important orchestration challenge at SaaS layer is: how does a SaaS offering compare to other relevant offerings? For example, how does a user compare the QoS and service features of SalesFoces.com, NetSuite, Clarizen.com, and Appirio? Ideally, the SaaS provider should provide guarantees on QoS as part of SLA, as well as outline clear policies and guidelines for SaaS maintenance and upgrades to simplify selection.

**Data Mining and Aggregation.** Social networking media and blogs (e.g. Facebook, Wordpress, MySpace, Blogspot, LinkedIn, Digg.com, etc.) generates massive amount of contents, which remains largely untapped. Recent studies have shown that critical analysis of these contents can be selectively aggregated to generate important new knowledge for predicting real-world outcomes (e.g. USA 2011 debt crisis, movie box-office review, etc.) and providing information on emergency situations to crisis management centers. Recently, HP Research Lab has demonstrated how the tweets from Twitter can be used for forecasting the box-office revenues of movies. The analogy here is that the rate at which contents are created by social networking media and blogs about particular topics can outperform existing market-based predictors and news media services. However, there are certain orchestration challenges in building such aggregated, real-time information services including how to efficiently process high volume of streaming data, how to search and query (data mining) text patterns, and how to aggregate disparate data sources that may have different API and data representation formats.

**Location-aware Network QoS Optimization.** Users of online multimedia content (data) services (e.g. Youtube) tend to follow certain download and streaming behaviors based on their own geo-distribution, the popularity and type of the content. It is possible to provide the further optimization and QoS guarantees in delivery of multimedia content, if the resource orchestrator is aware of the download behavior of end-users. For example, to achieve high network performance, an orchestrator can migrate the multimedia contents to the cloud, which is topologically most optimized to serve the end-users. Since IaaS provider are building data centers across the globe, therefore achieving such adaptive migration of content is not a difficult undertaking. However, the key challenges exist in detecting end-users network hotspots and initiating dynamic migration of contents.

**Data Protability.** SaaS provider (such as SalesForce.com, NetSuite) need to provide service features that allow users (e.g., SMEs, governments, etc.) to migrate their existing application data to cloud environments with minimal effort. All existing in-house application data need to be transparently exported, formatted, and parsed to suit the data format supported by cloud-hosted SaaS installation.

**Data Security and Privacy.** SaaS installations maintain confidential users data (e.g., emails, customer record, inventory reports, office documents). In particular, Identity management: authentication and authorization of SaaS users; provisioning end-user access; single-sign on, federated security model. Issues related to data confidentiality, integrity and availability discussed in Section 2.1; is also applicable here.

## 4. Conclusion and Related Works

Cloud computing is a vast, complex, and evolving technology landscape, embracing multi-layered resource stack (IaaS, PaaS, and SaaS) that must be orchestrated in an intricate manner to ensure that application delivers acceptable QoS level to the end-users. We characterized of cloud resources in a multi-layered stack based on their attributes, granularity, and supported orchestration operations. The paper will help readers in clearly understanding the core cloud computing concepts, inter-relationship between different resource types, and relevant research challenges. This in turn may lead to a harmonization of research efforts and more inter-operable cloud technologies.

Few recent papers [79][80][81][82][83][84][85] attempt to study the taxonomy of services in cloud computing landscape. The authors in the paper [37] proposed unified ontology for describing the cloud resource. They use composability as the methodology for developing the resource ontology. Although these papers help in understanding the issues of cloud computing, they fail to capture the essence of programming resource orchestration frameworks and associated research challenges. Further, they did not considered all resource types that we captured in this paper. Arguably, this paper is the first attempt at capturing orchestration operations and related research issues involved with programming orchestration framework across all layers of cloud computing resource stack (IaaS, PaaS, and SaaS).